\documentstyle[preprint,aps]{revtex}
\input epsfig.sty
\tightenlines
\begin{document}
\title{Anomalous Capacitive Sheath with Deep Radio Frequency Electric Field
Penetration}
\author{Igor D. Kaganovich,}
\address{Plasma Physics Laboratory, Princeton University, Princeton, NJ 08543%
}
\date{\today}
\maketitle

\begin{abstract}
A novel nonlinear effect of anomalously deep penetration of an
external radio frequency electric field into a plasma is
described. A self-consistent kinetic treatment reveals a
transition region between the sheath and the plasma. Because of
the electron velocity modulation in the sheath, bunches in the
energetic electron density are formed in the transition region
adjusted to the sheath. The width of the region is of order
$V_{T}/\omega $, where $V_{T}$ is the electron thermal velocity,
and $\omega $\ is frequency of the electric field. The presence of
the electric field in the transition region results in a cooling
of the energetic electrons and an additional heating of the cold
electrons in comparison with the case when the transition region
is neglected.

 \bigskip
\end{abstract}

The penetration of the electric field perpendicular to the plasma boundary
was studied by Landau in the linear approximation \cite{Landau}. He showed
that an external electric field with amplitude $E_{0}$ is screened by the
plasma electrons in the sheath region in a distance of order the Debye
length, and reaches a value $E_{0}/\varepsilon $ in the plasma, where $%
\varepsilon ${\small \ }is plasma dielectric constant. In many practical
applications, the value of the external electric field is large: the
potential drop in the sheath region $V_{sh}$ is typically of order hundreds
of Volts and is much larger than electron temperature $T_{e}$, which is of
order of a few Volts; and the field penetration has to be treated
nonlinearly. The asymptotic solution of sheath structure has been studied by
Lieberman in the limit $V_{sh}>>T_{e}$ \cite{Libermann89}. In this
treatment, the plasma sheath boundary is considered to be infinitely thin
and the position of the boundary is determined by the condition that the
external electric field is screened in the sheath regions when electrons are
absent. Electron interactions with the sheath electric field are
traditionally treated as collisions with a moving potential barrier (wall).
It is well known that multiple electron collisions with an oscillating wall
result in electron heating, provided there is sufficient phase-space
randomization in the plasma bulk. It is common to describe the sheath
heating by considering the electrons as test particles, and neglecting the
plasma electric fields \cite{Lieberman& Godyak review}. Kaganovich and
Tsendin proved in Ref.\cite{Me 1992} that accounting for the electric field
in the plasma reduces the electron sheath heating, and the electron sheath
heating vanishes completely in the limit of uniform plasma density.
Therefore, an accurate description of the rf fields in the bulk of the
plasma is necessary for calculating the sheath heating. The electron
velocity is oscillatory in the sheath, and as a result of this velocity
modulation electron density bunches appear in the region adjusted to the
sheath. The electron density perturbations decay due to phase mixing over a
length of order $V_{T}/\omega ,$ where $V_{T}$ is the electron thermal
velocity, and $\omega $\ is the frequency of the electric field. The
electron density perturbations polarize the plasma and produce an electric
field in the plasma bulk. This electric field, in turn, changes the velocity
modulation and correspondingly influences the electron density
perturbations. Therefore, electron sheath heating has to be studied in a
self-consistent nonlocal manner assuming a finite temperature plasma.

Notwithstanding the fact, that particle-in-cell simulations results are
widely available for the past decade [5-7] a basic understanding of the
electron sheath heating is incomplete, because no one has studied the
electric field in the plasma bulk using a nonlocal approach, similar to the
anomalous skin effect for inductive electric field \cite{Lifshitz}. In this
regard, analytical models are of great importance because they shed light on
the most complicated features of collisionless electron interactions with
the sheath. In this Letter, an analytical model is developed to explore the
effects associated with the self-consistent non-local nature of the
phenomenon.

One of the approaches to study electron sheath heating is based on a fluid
description of the electron dynamics. For the collisionless case, closure
assumptions for the viscosity and heat fluxes are necessary. In most cases,
the closure assumptions are made empirically or phenomenologically [6, 7].
The closure assumptions have to be justified by direct comparison with the
results of kinetic calculations as is done, for example, in Ref. \cite%
{Hammett}. Otherwise, inaccurate closure assumptions may lead to misleading
results as discussed below.

To model the sheath-plasma interaction analytically, the following
simplifying assumptions have been adopted. The discharge frequency is
assumed to be small compared with the electron plasma frequency. Therefore,
most of the external electric field is screened in the sheath region by an
ion space charge. The ion response time is typically larger than the inverse
discharge frequency, and the ion density profile is quasi-stationary. There
is an ion flow from plasma bulk towards electrodes. In the sheath region,
ions are being accelerated towards the electrode by the large sheath
electric field, and, the ion density in the sheath region is small compared
with the bulk ion density. In the present treatment, the ion density profile
is assumed fixed and is modeled in a two-step approximation: the ion density
$n_{b}$ is uniform in the plasma bulk, and the ion density in the sheath $%
n_{sh}<n_{b}$ is also uniform (see Fig.1). At the sheath-plasma boundary,
there is a stationary potential barrier for the electrons ($e\Phi _{sh}$),
so that only the energetic electrons reach the sheath region. The potential
barrier is determined by the quasineutrality condition, i.e., when the
energetic electrons enter the sheath region, their mean density is equal to
the ion density [$n_{e}(\Phi _{sh})=n_{sh}$].

The electron density profile is time-dependent in response to the
time-varying sheath electric field. The large sheath electric field does not
penetrate into the plasma bulk. Therefore, the quasineutrality condition
holds in the plasma bulk, i.e., the electron density is equal to ion
density, $n_{e}=n_{b}.$ In the sheath region, the electrons are reflected by
the large sheath electric field. Therefore, $n_{e}=n_{sh}$ for $x>x_{sh}(t)$%
, and $n_{e}=0$ for $x<x_{sh}(t)$,\ where $x_{sh}(t)$ is the position of the
plasma-sheath boundary \cite{Libermann89}. From Maxwell's equations it
follows that ${\bf \nabla \cdot J}=0$, where the total current ${\bf J}$ is
the sum of the displacement current and the electron current. In the
one-dimensional case, the condition ${\bf \nabla \cdot J}=0$ yields the
conservation of the total current:
\begin{equation}
en_{e}V_{e}+\frac{1}{4\pi }\frac{\partial E_{sh}}{\partial t}=j_{0}\sin
(\omega t+\phi ),  \label{Esh}
\end{equation}%
where $j_{0}$ is the amplitude of the rf current controlled by the external
circuit and $\phi $ is the initial phase. In the sheath, electrons are
absent in the region of large electric field, and the Eq.(\ref{Esh}) can be
integrated to give \cite{Me 1992}

\begin{equation}
E_{sh}(x,t)=\frac{4\pi j_{0}}{\omega }[-1-\cos (\omega t+\phi )]+4\pi
|e|n_{sh}x,\quad x<x_{sh}(t)  \label{Esh(x,t)}
\end{equation}%
where Poisson's equation has been used to determined the spatial dependence
of the sheath electric field. The first term on the right-hand side of Eq.(%
\ref{Esh(x,t)}) describes the electric field at the electrode, the second
term relates to ion space charge screening of the sheath electric field. The
position of the plasma-sheath boundary $x_{sh}(t)$ is determined by the zero
of the sheath electric field, $E_{sh}[x_{sh}(t),t]=0$. From Eq.(\ref%
{Esh(x,t)}) it follows that
\begin{equation}
x_{sh}(t)=\frac{V_{sh0}}{\omega }[1+\cos (\omega t+\phi )],  \label{xsh(t)}
\end{equation}%
where $V_{sh0}=j_{0}/(en_{sh})$ is the amplitude of the plasma-sheath
boundary velocity. The ion flux on the electrode is small compared with the
electron thermal flux. Because electrons attach to the electrode, the
electrode surface charges negatively, so that in a steady-state discharge,
the electric field at the electrode is always negative, preventing an
electron flux on the electrode. However, for a very short time ($\omega
t_{n}+\phi \approx \pi (1+2n)$) the sheath electric field vanishes, allowing
electrons to flow to the electrode for compensation of the ion flux. Note
that there is a large difference between the sheath structure in the
discharge and the sheath for obliquely incident waves interacting with a
plasma slab without any bounding walls. Because electrodes are absent,
electrons can move outside the plasma, and the electric field in the vacuum
region, $E_{sh}(x,t)=(4\pi j_{0}/\omega )\cos (\omega t+\phi )$, may have a
different sign. Therefore, electrons may penetrate into the region of large
electric field during time when $E_{sh}(x,t)<0$ [10,11]. However, in the
discharge, because the sheath electric field given by Eq.(\ref{Esh(x,t)}) is
always reflecting electrons, the electrons {\em never} enter the region of
the large sheath electric field, which is opposite to the case of obliquely
incident waves.

The calculations based on the two-step ion density profile model is known to
yield discharge characteristics in good agreement with experimental data and
full-scale simulations \cite{Orlov}.

Throughout this paper, linear theory is used because the plasma-sheath
boundary velocity and the mean electron flow velocity are small compared
with the electron thermal velocity [4,5]. The important spatial scale is the
length scale for phase mixing, $l_{mix}=V_{T}/\omega $. The sheath width
satisfies $2V_{sh0}/\omega <<l_{mix}$ because $V_{sh}<<V_{T}$. Therefore,
the sheath width is neglected, and electron interactions with the sheath
electric field are treated as a boundary condition. The collision frequency (%
$\nu $) is assumed to be much less than the discharge frequency ($\nu
<<\omega $), and correspondingly the mean free path is much larger than the
length scale for phase mixing. Therefore, the electron dynamics is assumed
to be collisionless. The discharge gap is considered to be sufficiently
large compared with the electron mean free path, so that the influence of
the opposite sheath is neglected. The effects of finite gap width are
discussed in Ref. \cite{Me PRL 1999}.

The electron interaction with the large electric field in the sheath is
modelled as collisions with a moving oscillating rigid barrier with velocity
$V_{sh}(t)=dx_{sh}(t)/dt$. An electron with initial velocity $-u$ after a
collision with the plasma-sheath boundary - modeled as a rigid barrier
moving with velocity $V_{sh}(t)$ - acquires a velocity $u+2V_{sh}$.
Therefore, the power deposition density transfer from the oscillating
plasma-sheath boundary is given by \cite{Libermann89}

\begin{equation}
P_{sh}=\frac{m}{2}\left\langle \int_{-V_{sh}}^{\infty }du\left[ u+V_{sh}(t)%
\right] \left[ (2V_{sh}(t)+u)^{2}-u^{2}\right] \,f_{sh}(-u,t)\right\rangle ,
\label{Psh}
\end{equation}%
where $m$ is the electron mass, $f_{sh}(-u,t)$ is the electron velocity
distribution function in the sheath, and $\left\langle \cdot \cdot \cdot
\right\rangle $ denotes a time average over the discharge period.
Introducing a new velocity distribution function $g(-u^{\prime
},t)=f_{sh}[-u-V_{sh}(t),t]$, Eq.(\ref{Psh}) yields

\begin{equation}
P_{sh}=-2m\left\langle V_{sh}(t)\int_{0}^{\infty }u^{\prime 2}g(-u^{\prime
},t)du^{\prime }\right\rangle ,  \label{Psh modified}
\end{equation}%
where $-u^{\prime }=-u-V_{sh}$ is the electron velocity relative to the
oscillating rigid barrier. From Eq.(\ref{Psh modified}) it follows that, if
the function $g(u^{\prime })$ is stationary, then ($P_{sh}=0$) there is no
collisionless power deposition due to electron interaction with the sheath
[7, 14]. For example, in the limit of a uniform ion density profile $%
n_{sh}=n_{b}$, $g(u^{\prime })$ is stationary ({\em in an oscillating
reference frame of the plasma-sheath boundary}), and the electron heating
vanishes \cite{Me 1992}. Indeed, in the plasma bulk the displacement current
is small compared with the electron current, and from Eq.(\ref{Esh}) it
follows that the electron mean flow velocity in the plasma bulk, $%
V_{b}(t)=-j_{0}\sin (\omega t+\phi )/|e|n_{b}$, is equal to the
plasma-sheath velocity $V_{sh}(t)$, from Eq.(\ref{xsh(t)}). Therefore, the
electron motion in the plasma is strongly correlated with the plasma-sheath
boundary motion. From the electron momentum equation it follows that there
is an electric field, $E_{b}=m/e\,dV_{b}(t)/dt$, in the plasma bulk.\ In a
frame of reference moving with the electron mean flow velocity, the sheath
barrier is stationary, and there is no force acting on the electrons,
because the electric field is compensated by the inertial force ($%
eE_{b}-mdV_{b}(t)/dt=0)$. Therefore, electron interaction with the sheath
electric field is totally compensated by the influence of the bulk electric
field, and the collisionless heating vanishes \cite{Me 1992}.

The example of a uniform density profile shows the importance of a
self-consistent treatment of the collisionless heating in the plasma. If the
function $g(u^{\prime },t)$ is nonstationary, there is net power deposition.
In this Letter, a kinetic calculation is performed to yield the correct
electron velocity distribution function $g(u^{\prime },t)$ and,
correspondingly, the net power deposition.

The electron motion is different for the low energy electrons with initial
velocity in the plasma bulk $|u|<u_{sh}$, where $u_{sh}^{2}=2e\Phi _{sh}/m$%
,\ and for the energetic electrons with velocity $|u|>u_{sh}$. The low
energy electrons with initial velocity in the plasma bulk $-u$ are reflected
from the stationary potential barrier $e\Phi _{sh}$, and then return to the
plasma bulk with velocity $u$. High energy electrons enter the sheath region
with velocity $u_{1}=-(u^{2}-u_{sh}^{2})^{1/2}$. They have velocity $%
u_{2}=2V_{sh}-u_{1}$ colliding with the moving rigid barrier,\ and then
return to the plasma bulk with velocity $(u_{2}^{2}+u_{sh}^{2})^{1/2}$ \cite%
{multiple collisions}.

As the electron velocity is modulated in time during reflections from the
plasma-sheath boundary, so is the energetic electron density (by continuity
of electron flux). This phenomenon is identical to the mechanism for
klystron operation \cite{klystron}. The perturbations in the energetic
electron density yield an electric field in the transition region adjusted
to the sheath.

The electron velocity distribution function is taken to be a sum of a
stationary isotropic part $f_{0}(u)$ and a nonstationary anisotropic part $%
f_{1}(x,u,t)$. $f_{1}$ is to be of the form $f_{1}(x,u,t)=f_{1}(x,u)\exp
(-i\omega t)$. The linearized Vlasov equation becomes

\begin{equation}
-i\omega f_{1}+u\frac{\partial f_{1}}{\partial x}+\frac{eE(x)}{m}\frac{df_{0}%
}{du}=-\nu f_{1},  \label{Vlasov eq.}
\end{equation}%
where the term on the right-hand side accounts for rare collisions ($\nu
<<\omega $). All time-dependent variables are assumed to be harmonic
functions of time, proportional to $\exp (-i\omega t),$ and, in the
subsequent analysis, the multiplicative factor $\exp (-i\omega t)$ is
omitted from the equations. The electron velocity distribution function must
satisfy the boundary condition at the plasma-sheath boundary ($x=0$)
corresponding to $f(0,u)=f(0,-u)$ for
\mbox{$\vert$}%
$u|<u_{sh}$,\ and $f_{sh}(u^{\prime })=f_{sh}(2V_{sh}-u^{\prime }),$ for $%
u>u_{sh}$, where $u^{\prime }=(u^{2}-u_{sh}^{2})^{1/2}$ \ and $f_{sh}$\ is
the electron velocity distribution in the sheath. From energy and flux
conservation, $u^{\prime }f_{sh}(u^{\prime })du^{\prime }=uf(u)du$,\ it
follows that $f_{sh}(u^{\prime })=f[(u^{\prime 2}+u_{sh}^{2})^{1/2}]$.\
Linearly approximating the boundary conditions yields

\begin{equation}
f_{1}(0,u)=f_{1}(0,-u),\,\quad 0<u<u_{sh},  \label{b.c.a)}
\end{equation}%
\begin{equation}
f_{1}(0,u)=f_{1}(0,-u)+2V_{sh}\frac{u^{\prime }}{u}\frac{df_{0}}{du}\,,\quad
u>u_{sh}.  \label{b.c. bb)}
\end{equation}%
The electric field is determined from the condition of conservation of the
total current ($j_{0}$), which gives
\begin{equation}
e\int_{-\infty }^{\infty }uf_{1}(x,u)du-\frac{i\omega }{4\pi }E(x)=j,
\label{Eq. for E}
\end{equation}%
where $j=\,j_{0}e^{i(\phi +\pi /2)}$, and the first term is the electron
current and the second term corresponds to a small displacement current.
Equations (\ref{Vlasov eq.}) and (\ref{Eq. for E}), together with the
boundary conditions (\ref{b.c.a)}), (\ref{b.c. bb)}) comprise the full
system of equations for the bulk plasma.

It is convenient to solve Eq. (\ref{Vlasov eq.}) by continuation into the
region $x<0$. First, we introduce the artificial force
\begin{equation}
F(x,u)=2mV_{sh}u^{\prime }\delta (x)\Theta (|u|-u_{sh}),
\label{Artificial F}
\end{equation}%
where $V_{sh}=j/en_{sh}$,  $\delta (x)$ is the Dirac delta-function, and $%
\Theta (u)$ is the Heaviside step function. The force in Eq.(\ref{Artificial
F}) accounts for the change of the energetic electron velocity in the sheath
region. Equation (\ref{Vlasov eq.}) together with the boundary conditions (%
\ref{b.c.a)}) and (\ref{b.c. bb)}) are equivalent to Eq. (\ref{Vlasov eq.})
with the force in Eq.(\ref{Artificial F}) added to the third term of Eq. (%
\ref{Vlasov eq.}). This gives

\begin{equation}
-i\omega f_{1}+u\frac{\partial f_{1}}{\partial x}+\frac{eE(x)+F(x,u)}{m}%
\frac{df_{0}}{du}=-\nu f_{1},  \label{Vlasov eq. modified}
\end{equation}%
where the boundary condition (\ref{b.c.a)}) for all electrons becomes%
\begin{equation}
f_{1}(0,u)=f_{1}(0,-u).  \label{b.c. all}
\end{equation}

In this formulation, the half-space problem is equivalent to that of an
infinite medium in which the electric field is antisymmetric about the plane
$x=0$, with $E(x)=-E(-x)$ [1, 17]. Such a continuation makes Eq. (\ref%
{Vlasov eq. modified}) invariant with respect to the transformation $%
x\rightarrow -x$, and $u\rightarrow -u$. Electrons reflected from the
boundary in the half-space ($x>0$) problem correspond to electrons passing
freely through the plane $x=0$ from the side $x<0$ in the infinite-medium
problem.

A spatial Fourier the transform of Eq. (\ref{Vlasov eq. modified}) gives%
\begin{equation}
f_{1}(k)=\frac{eE(k)+F_{sh}(u)}{mi(\omega -uk+i\nu )}\frac{df_{0}}{du},
\label{Fourier of f1}
\end{equation}%
where $E(k)$\ is the Fourier transform of $E(x)$

\begin{equation}
E(k)=\int_{-\infty }^{\infty }E(x)\exp (-ikx)dx,  \label{E(k)}
\end{equation}%
and $F_{sh}(u)=2mV_{sh}u^{\prime }\Theta (|u|-u_{sh}).$ It is convenient to
divide the electric field in the plasma into two parts corresponding to $%
E(x)=E_{1}(x)+E_{b}sgn(x)$, where $E_{1}(x)\rightarrow 0$ for $x\rightarrow
\infty $, and $E_{b}$ is the value of the electric field far away from the
sheath region. The Fourier transform of the electron current can be obtained
by integrating Eq. (\ref{Fourier of f1}) over velocity, yielding%
\begin{equation}
j(k)=\sigma (k)E_{1}(k)-\frac{2i}{k}\left[ E_{sh}\sigma _{sh}(k)+\sigma
(k)E_{b}\right] ,  \label{j(k)}
\end{equation}%
\begin{equation}
\sigma (k)=-\frac{ie^{2}}{m}\int_{-\infty }^{\infty }\frac{u}{(\omega
-uk+i\nu )}\frac{df_{0}}{du}du,  \label{s(k)}
\end{equation}%
\begin{equation}
\sigma _{sh}(k)=\frac{ie^{2}k}{(\omega +i\nu )m}\int_{-\infty }^{\infty }%
\frac{uu^{\prime }\Theta (|u|-u_{sh})}{(\omega -uk+i\nu )}\frac{df_{0}}{du}%
du,  \label{nsh(k)}
\end{equation}%
where $\sigma (k)$ is the electron conductivity, $\sigma _{sh}(k)$ is the
effective conductivity due to electron interaction with the sheath, and $%
E_{sh}=(-i\omega +\nu )mV_{sh}/e$ is the effective electric field
corresponding to $V_{sh}$.

The Fourier amplitude $E_{1}(k)$ is to be determined from Eq.(\ref{Eq. for E}%
) continued into the half-space $x<0$. Because $E(x)$ is an antisymmetric
function about the plane $x=0$, $j_{0}$ is continued with negative sign into
the half-space $x<0$, and the Fourier transform of $j_{0}sgn(x)$ is $%
-2i\,j_{0}/k$. Substituting $E(k)=E_{1}(k)-2iE_{b}/k$ and $j_{0}=\left[
\sigma (0)-i\omega /4\pi \right] E_{b}$\ into Fourier transform of Eq.(\ref%
{Eq. for E}) gives
\begin{equation}
E_{1}(k)=-\frac{2i}{k}\frac{[\sigma (0)-\sigma (k)]E_{b}-E_{sh}\sigma
_{sh}(k)}{\sigma (k)-\frac{i\omega }{4\pi }}.
\label{Eg. for E1 in F. transform}
\end{equation}%
Notice that, if the plasma density in the sheath is equal to the bulk
density $n_{sh}=n_{b}$, then $u_{sh}=0$, $E_{b}=E_{sh}$ and $\sigma
(0)-\sigma (k)=\sigma _{sh}(k)$. Therefore, $E_{1}(k)=0$ and the uniform
electric field $E_{b}$ satisfies the current conservation condition, as
discussed earlier.

The profile for $E_{1}(x)$ given by inverse Fourier transform
\begin{equation}
E(x)=\frac{1}{2\pi }\int_{-\infty }^{\infty }E(k)\exp (ikx)dk  \label{E(x)}
\end{equation}%
is shown at the top in Fig.2. For $x<6V_{T}/\,\omega $ the electric field
profile is close to $E_{1}(x)\approx E_{1}(0)\exp (-\lambda x\omega /V_{T}\,)
$, where $E_{1}(0)=-0.72$, and $\lambda =0.19+0.77i$ for the conditions in
Fig.2. For $x>6V_{T}/\,\omega $,\ the electric field profile is no longer a
simple exponential function, similar to the case of the anomalous skin
effect \cite{Aliev and me}. The three components of current corresponding to
the first, second, and third terms in Eq. (\ref{j(k)}) are shown at the
bottom in Fig.2.\ The first term describes the current ($j_{tr}$) driven by
the electric field $E_{1}(x)$ under the assumption of specular reflection at
the boundary. The second term relates the current ($j_{sh}$) of the
energetic electrons owing only to a velocity change due to reflections from
the large sheath electric field. The third term describes the current ($j_{b}
$) driven by the uniform electric field $E_{b}$ under the assumption of
specular reflection at the boundary. Due to the boundary condition of
specular reflection in Eq. (\ref{b.c.a)}), both of the currents $j_{b}$ and $%
j_{tr}$ are equal to zero at $x=0$. Also, both of the currents $j_{tr}$ and $%
j_{sh}$ vanish at $x>15V_{T}/\,\omega $ due to phase mixing, and the only
current left here is $j_{b}.$ In contrast to large $x$, at small $%
x<<V_{T}/\,\omega $ the total current is entirely due to energetic electrons
interacting with the sheath $j_{sh}$. Indeed, the energetic electrons enter
the sheath region with velocity distribution $f_{sh}(u^{\prime })$. The
electron current is given by the sum of the contribution from the electrons
approaching the oscillating barrier and from the electrons already reflected
from the barrier, $j_{sh}=\int_{V_{sh}}^{\infty }u^{\prime }f_{sh}(u^{\prime
})du^{\prime }+\int_{-\infty }^{-V_{sh}}u^{\prime }f_{sh}(u^{\prime
})du^{\prime }.$ Because $f_{sh}(u^{\prime })=f_{sh}(2V_{sh}-u^{\prime })$,
\ $j_{sh}=2eV_{sh}\int_{V_{sh}}^{\infty }f_{sh}(u)du\approx
eV_{sh}\int_{-\infty }^{\infty }f_{sh}(u^{\prime })du^{\prime
}=en_{sh}V_{sh}=j_{o}\sin (\omega t+\phi ).$ In the last calculation the
contribution to the density by electrons with velocity $u<V_{sh0}$ is
omitted. Their contributions are second-order effects in $V_{sh0}/V_{T}$,
which are neglected in the present study \cite{multiple collisions}.
Therefore, in the sheath region, when electrons are present, and in the
nearest vicinity of the sheath all current is conducted by the energetic
electrons. As can be seen in Fig.2, the current conservation condition, $%
j_{tr}(x)+j_{sh}(x)+j_{b}(x)=\left[ \sigma (0)-i\omega /4\pi \right] E_{b}$,
is satisfied for arbitrary $x$.

The difference in phase of the currents of the energetic and low energy
electrons was observed in Ref.\cite{Surendra PRL}, but it was misinterpreted
as the generation of electron acoustic waves. Electron acoustic waves can be
excited if the denominator of the right-hand side of Eq. (\ref{Eg. for E1 in
F. transform}) has a pole at frequency $\omega $, which corresponds to the
root of the plasma dielectric function, $\varepsilon =1+4\pi i\sigma
(k)/\omega $. For a Maxwellian electron distribution function, the pole does
not exist for $\omega <<\omega _{p}$, where $\omega _{p}=\sqrt{4\pi
e^{2}n_{b}/m}$ is the electron plasma frequency. But the electron acoustic
waves can exist if the plasma contains two groups of electrons having very
different temperatures \cite{Mace}. The wave phase velocity is $\omega /k=%
\sqrt{n_{c}/n_{h}}\sqrt{T_{h}/m}$ , where $n_{c}$ and $n_{h}$ are the
electron density of cold and hot electrons, respectively, and $T_{h}$ is the
temperature of the hot electrons. The electron acoustic waves are strongly
damped by the hot electrons, unless $n_{c}<<n_{h}$ and $T_{c}<<T_{h}$ ,
where $T_{c}$ is the electron temperature of the cold electrons \cite{Mace}.
In the opposite limit, $n_{c}>4n_{h}$, the electron acoustic waves do not
exist \cite{Mace}. In capacitively-coupled discharges, the electron
population does stratify into two populations of cold and hot electrons, as
has been observed in experiments and simulation studies [19,20]. Cold
electrons trapped in the discharge center by the plasma potential do not
interact with the large electric fields in the sheath region and have low
temperature. Moreover, because of the nonlinear evolution of plasma
profiles, the cold electron density is much larger than the hot electron
density \cite{Me 2EDF}. Therefore, weakly-damped electron acoustic waves do
not exist in the plasma of capacitively-coupled discharges. Reference \cite%
{Surendra PRL}\ used the fluid equation and neglected the effect of
collisionless dissipation, thus arriving at the wrong conclusion about the
existence of weakly-damped electron acoustic waves.

The power deposition is given by the sum of the power transferred to the
electrons by the oscillating rigid barrier in the sheath region and by the
electric field in the transition region,%
\begin{equation}
P_{tot}=P_{sh}+P_{tr}.  \label{Power tot}
\end{equation}%
Here $P_{sh}$ is given by Eq.(\ref{Psh}), which after linearization yields%
\begin{equation}
P_{sh}=P_{sh0}+P_{sh1.}  \label{Psh=Psh0+1}
\end{equation}%
In Eq.(\ref{Psh=Psh0+1}), $P_{sh0}$ is the power dissipation in the sheath
neglecting any influence of electric field,

\begin{equation}
P_{sh0}=2m\left\langle \int_{0}^{\infty }2u^{\prime
}\,V_{sh}(t)^{2}\,f_{0sh}(-u^{\prime },t)\right\rangle ,  \label{Psh0}
\end{equation}%
and $P_{sh1}$ accounts for the influence of the electric field on $f_{1}$
and correspondingly on the power dissipation in the sheath,
\begin{equation}
P_{sh1}=2m\left\langle \int_{0}^{\infty }V_{sh}(t)u^{\prime
2}\,f_{1sh}(-u^{\prime },x=0,t)du^{\prime }\right\rangle .  \label{Psh1}
\end{equation}%
Time averaging, changing variables from $u^{\prime }$ to $u$, and
integration by parts in the first term yield%
\begin{equation}
P_{sh}=m\int_{0}^{\infty }\left\{ -|V_{sh}|^{2}\,u^{\prime 2}\,\frac{df_{0}}{%
du}+%
\mathop{\rm Re}%
\left[ V_{sh}u^{\prime }\,uf_{1}^{\ast }(-u,x=0)\right] \right\} \Theta
(|u|-u_{sh})du,  \label{Psh2}
\end{equation}%
where $f_{1}^{\ast }$ is solution to Eq.(\ref{Vlasov eq.}),%
\begin{equation}
f_{1}^{\ast }(-u,x=0)=\frac{e}{mu}\frac{df_{0}}{du}\int_{0}^{\infty }E^{\ast
}(x)e^{-(i\omega +\nu )x/u}dx.  \label{f1*}
\end{equation}

Time averaging the power deposition in the transition region, $%
\int_{0}^{\infty }\left\langle jE\right\rangle dx$, gives%
\begin{equation}
P_{tr}=\frac{1}{2}%
\mathop{\rm Re}%
\int_{0}^{\infty }j_{0}E^{\ast }dx.  \label{P_E}
\end{equation}%
Substituting $j_{0}=ien_{b}V_{b}$, where $V_{b}=eE_{b}/m\omega $ is the
amplitude of the mean electron flow velocity in the plasma bulk and $\phi =0$
was assumed in Eq.(\ref{Esh}), we obtain\ $P_{b}=1/2%
\mathop{\rm Re}%
j_{0}E^{\ast }=-1/2en_{b}V_{b}%
\mathop{\rm Im}%
E_{1}(x)$. Therefore, $P_{b}$ is determined by the imaginary part of $E_{1}$%
, and can be either positive or negative (see Fig. 2). Negative power
density has been observed in numerical simulations \cite{Surendra PRL}.

Substituting $j_{0}=j_{E}+j_{sh}$, where $j_{E}=j_{b}+j_{tr}$, the power
deposited by the current $j_{E}$ can be calculated by continuing into
infinite space and using the Fourier transform \cite{Aliev and me}%
\begin{equation}
\frac{1}{2}%
\mathop{\rm Re}%
\int_{0}^{\infty }j_{E}E^{\ast }dx=\frac{1}{4}%
\mathop{\rm Re}%
\int_{-\infty }^{\infty }j_{E}E^{\ast }dx=\frac{1}{8\pi }%
\mathop{\rm Re}%
\int_{-\infty }^{\infty }j_{E}(k)E^{\ast }(k)dk,  \label{P_E2}
\end{equation}%
where $j_{E}(k)=\sigma (k)E(k).$ Finally, substituting the conductivity from
Eq.(\ref{s(k)}) yields
\begin{equation}
\frac{1}{2}%
\mathop{\rm Re}%
\int_{0}^{\infty }j_{E}E^{\ast }dx=-\frac{1}{4}\int_{0}^{\infty }\frac{%
e^{2}|E(k=\frac{\omega }{k})|^{2}}{m}\frac{df_{0}}{du}du.  \label{P_E3}
\end{equation}%
The current $j_{sh}$ is determined by the perturbed electron velocity
distribution function due to reflections from the sheath electric field. The
perturbed distribution function $f_{1sh}$ at $x=0$ is given by Eq.(\ref{b.c.
bb)}), and for $x>0$ the solution to the Vlasov equation becomes%
\begin{equation}
f_{1sh}(x,u)=-2V_{sh}\frac{u^{\prime }}{u}\frac{df_{0}}{du}e^{(i\omega -\nu
)x/u}.  \label{fsh}
\end{equation}%
Calculating the current $j_{sh}$ by integrating $f_{1sh}$ from Eq.(\ref{fsh}%
) over velocity, and substituting the current into Eq.(\ref{P_E}) gives%
\begin{equation}
\frac{1}{2}%
\mathop{\rm Re}%
\int_{0}^{\infty }j_{sh}E^{\ast }dx=-%
\mathop{\rm Re}%
\left[ V_{sh}\int_{0}^{\infty }u^{\prime }E^{\ast }(k=\frac{\omega }{u})%
\frac{df_{0}}{du}du\right] .  \label{P_Jsh}
\end{equation}%
Substituting $f_{1}^{\ast }$ from Eq. (\ref{f1*}) into Eq. (\ref{Psh2}), and
adding the contributions from Eqs.(\ref{P_E3}) and (\ref{P_Jsh}) yield%
\begin{equation}
P_{tot}=-\int_{0}^{\infty }muD_{u}(u)\frac{df_{0}}{du}du,  \label{Ptot}
\end{equation}%
where $D_{u}(u)$ is the diffusion coefficient in velocity space,
\begin{equation}
D_{u}(u)=\frac{u|du|^{2}}{4},  \label{D(u)}
\end{equation}%
and $du$ is the change in electron velocity after passing through the
transition and sheath regions,%
\begin{equation}
du=2iV_{b}\left[ \frac{u^{\prime }}{u}\frac{n_{b}}{n_{sh}}\Theta
(|u|-u_{sh})-1\right] +\frac{eE_{1}(k=\omega /u)}{u}.  \label{du}
\end{equation}%
A plot of $|du|^{2}/2$ is shown in Fig.3. Taking into account the electric
field in the plasma (both $E_{b}$ and $E_{1}$) reduces $|du|$ for energetic
electrons ($u>u_{sh}$) and increase $|du|$ for slow electrons ($u<u_{sh}$).
Therefore, the electric field in the the plasma cools the energetic
electrons and heats the low energy electrons, respectively. Similar
observations were made in numerical simulations \cite{Surendra PRL}.

Figure 4 shows the dimensionless power density as a function of $%
n_{b}/n_{sh} $. Taking into account the electric field in the plasma (both $%
E_{b}$ and $E_{1}$) reduces the total power deposited in the sheath region.
Interestingly, taking into account only the uniform electric field $E_{b}$
gives a result close to the case when both $E_{b}$ and $E_{1}$ are accounted
for. The electric field $E_{1}$ redistributes the power deposition from the
energetic electrons to the low energy electrons, but does not change the
total power deposition (compare Fig.3 and Fig.4). Therefore, the total power
deposition due to sheath heating can be calculated approximately from Eq. (%
\ref{Ptot}), taking into account only the electric field $E_{b}$. This gives%
\begin{equation}
P_{tot}\approx -mV_{b}^{2}\int_{0}^{\infty }u^{2}\left[ \frac{u^{\prime }}{u}%
\frac{n_{b}}{n_{sh}}\Theta (u-u_{sh})-1\right] ^{2}\frac{df_{0}}{du}du.\,
\label{Ptotfinal}
\end{equation}%
The result of the self-consistent calculation of the power dissipation in
Eq.(\ref{Ptotfinal}) differs from the non-self-consistent estimate in Eq.(%
\ref{Psh0}) by the last term in Eq.(\ref{Ptotfinal}), which contributes
corrections of order $n_{sh}/n_{b}$ to the main term.

This research was supported by the U.S. Department of Energy. The author
gratefully acknowledges helpful discussions with Ronald C. Davidson,
Vladimir I. Kolobov, Michael N. Shneider, Gennady Shvets, and Edward
Startsev.

\bigskip

{\Large APPENDIX:}

\section{Properties of $E_{1}(k)$}

The Fourier transform $E_{1}(k)$ has the following properties in the limits
of small and large $k$. At small $k$ ($k<<\omega /V_{T}$), $E_{1}(k)\sim k,$
because the numerator in the last factor on the right-hand side of Eq.(\ref%
{Eg. for E1 in F. transform})$\sim k^{2}$ ( [$\sigma (0)-\sigma (k)]\sim
k^{2}$ and $\sigma _{sh}(k)\sim k^{2}$). Because $E_{1}(k)\sim k$ for small $%
k$, $\int E_{1}(x)dx=0$ similarly to the case of anomalous skin effect \cite%
{Aliev and me}.

At large $k$ ($r_{d}^{-1}>>k>>\omega /V_{T}$), $E_{1}(k)\symbol{126}1/k$,
because both the numerator and the denominator in the last factor on the
right-hand side of Eq.(\ref{Eg. for E1 in F. transform}) are reciprocal to $%
k^{-2}$ ($\sigma (0)E_{b}=E_{sh}\sigma _{sh}(k\rightarrow \infty )$). $\
E_{1}(x)$ at small $x$ is determined by behavior of $E_{1}(k)$ at large $k.$
In the limit of large $k$ ($r_{d}^{-1}>>k>>\omega /V_{T}$)%
\begin{equation}
E_{1}(k)=\frac{2iA}{k},  \eqnum{A1}
\end{equation}%
where
\begin{equation}
A=E_{b}-\frac{C}{B}E_{sh}.  \eqnum{A2}
\end{equation}%
Here,
\begin{equation}
B=\lim_{k\rightarrow \infty }\sigma (k)k^{2},  \eqnum{A3}
\end{equation}%
\begin{equation}
C=\lim_{k\rightarrow \infty }\left[ \sigma _{sh}(\infty )-\sigma _{sh}(k)%
\right] k^{2}.  \eqnum{A4}
\end{equation}%
For a Maxwellian electron distribution function, substituting definitions of
conductivities Eqs.(\ref{s(k)}) and (\ref{nsh(k)}) into Eqs.(A3) and (A4),
respectively, yields%
\begin{equation}
B=\frac{ie^{2}\omega }{m}\int_{-\infty }^{\infty }\frac{1}{u}\frac{df_{0}}{du%
}du=\frac{-ie^{2}\omega n_{b}}{T}\   \eqnum{A5}
\end{equation}%
\begin{equation}
C=\frac{-ie^{2}\omega }{m}\int_{-\infty }^{\infty }\frac{u^{\prime }}{%
u^{\prime 2}+u_{sh}^{2}}\frac{df_{sh}}{du^{\prime }}du^{\prime }=\frac{%
-ie^{2}\omega n_{sh}}{T}\left( 1-\sqrt{\pi }u_{nsh}e^{u_{nsh}^{2}}\left[ 1-%
\mathop{\rm erf}%
(u_{nsh})\right] \right) ,  \eqnum{A6}
\end{equation}%
where $u_{nsh}=u_{sh}/V_{T},\ $and $%
\mathop{\rm erf}%
(u_{nsh})\ $is\ the error function. Form Eq.($\ref{E(x)}$),$\ E_{1}(x)$ at
small $x$ is given by$\ $%
\begin{equation}
E_{1}(x\rightarrow 0)=-\frac{2A}{\pi }\int_{0}^{\infty }\frac{\sin (kx)}{k}%
dk=-A.  \eqnum{A6}
\end{equation}%
Substituting and $E_{sh}=E_{b}n_{b}/n_{sh}$ and values of $B$ and $C$ from
Eqs. (A5) and (A6) into Eq.(A2) gives
\[
E_{1}(0)=-\sqrt{\pi }u_{nsh}e^{u_{nsh}^{2}}\left[ 1-%
\mathop{\rm erf}%
(u_{nsh})\right]
\]

\begin{figure}[h]
\epsfig{file=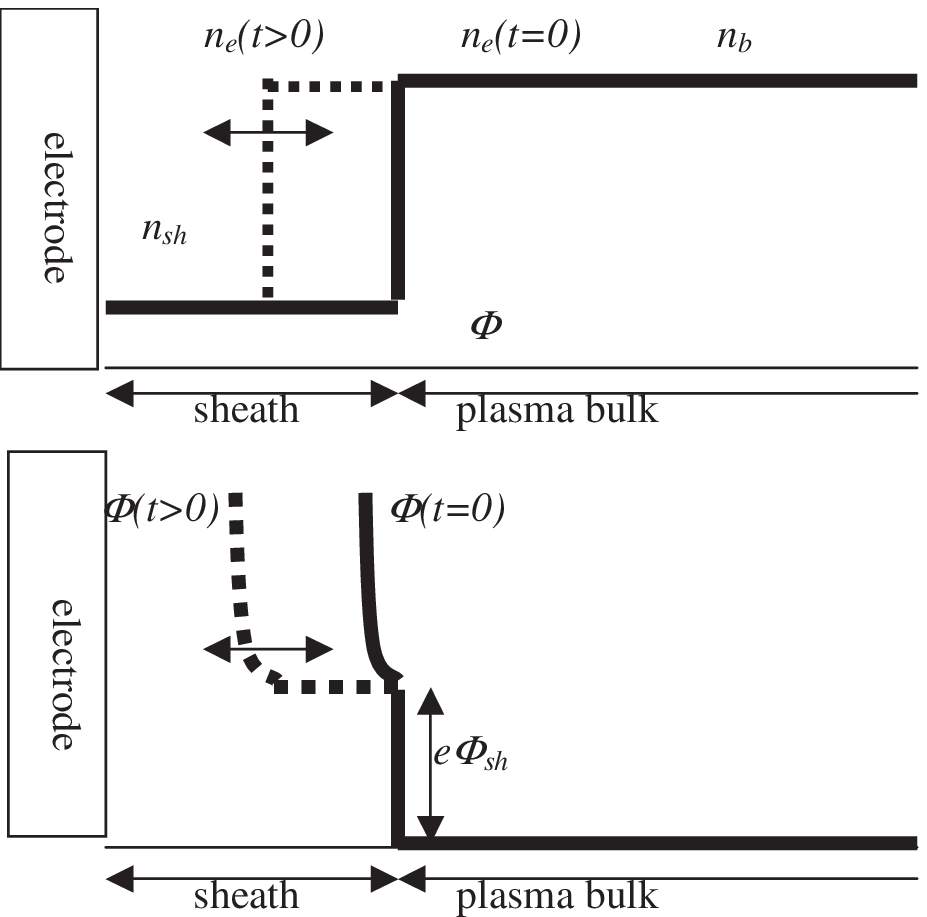}
 \caption{Schematic of a sheath. The negatively charged electrode pushes electrons away by different distances depending on the strength of the electric field at the electrode. Shown are the density and potential profiles at two different times. The solid line is at the time of maximum sheath expansion. }
\label{fig1}
\end{figure}

\begin{figure}[h]
\epsfig{file=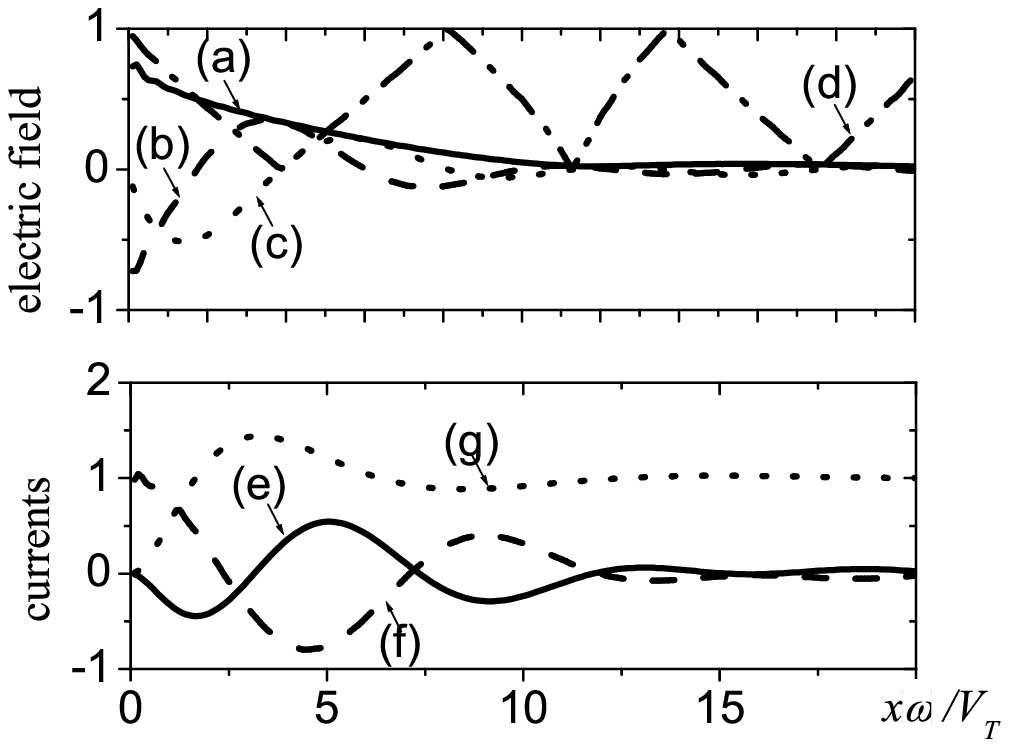}
 \caption{ Plots of the
electric field and the current normalized to their respective
values in the plasma bulk, $E_{b}$ and $e^{2}nE_{b}/m\omega $, as
functions of the normalized coordinate $x\omega /V_{T}$ for the
following parameters: $n_{sh}/n_{b}=1/3$, $\omega /\omega
_{p}=1/100$, and a Maxwellian electron distribution function. The
upper graph shows profiles of $E_{1}(x)$: (a) amplitude - solid
line; (b) real part - dashed line; (c) imaginary part - dotted
line; and (d) phase with respect to phase of $E_{b}$ divided by
$\pi $- dash-dotted line. The lower graph shows profiles of
imaginary part of currents: (e) $j_{tr}$ - solid line; (f)
$j_{sh}$ -dashed line; and (g) $j_{b}$ - dotted line.}
\label{fig2}
\end{figure}

\begin{figure}[h]
\epsfig{file=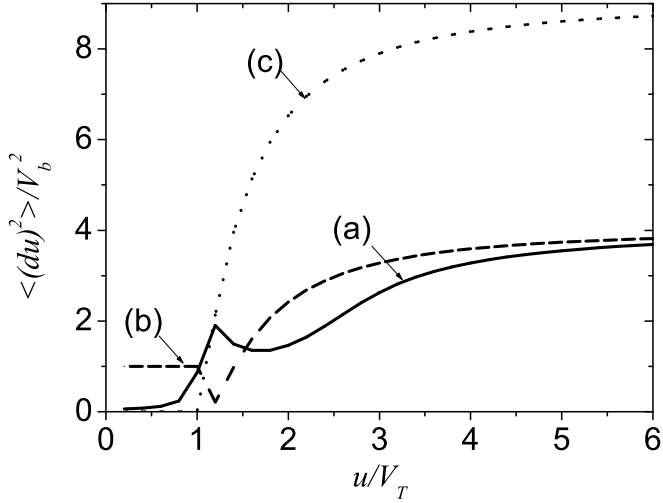}
\caption{ Plot of the average square of the
dimensionless velocity kick as a function of the dimensionless
velocity for the conditions in Fig.1, taking into account (a) both
$E_{1}(x)$ and $E_{b}$- solid line; (b) only $E_{b}$- dashed line;
and (c) no electric field - dotted line.}
 \label{fig3}
\end{figure}

\begin{figure}[h] \epsfig{file=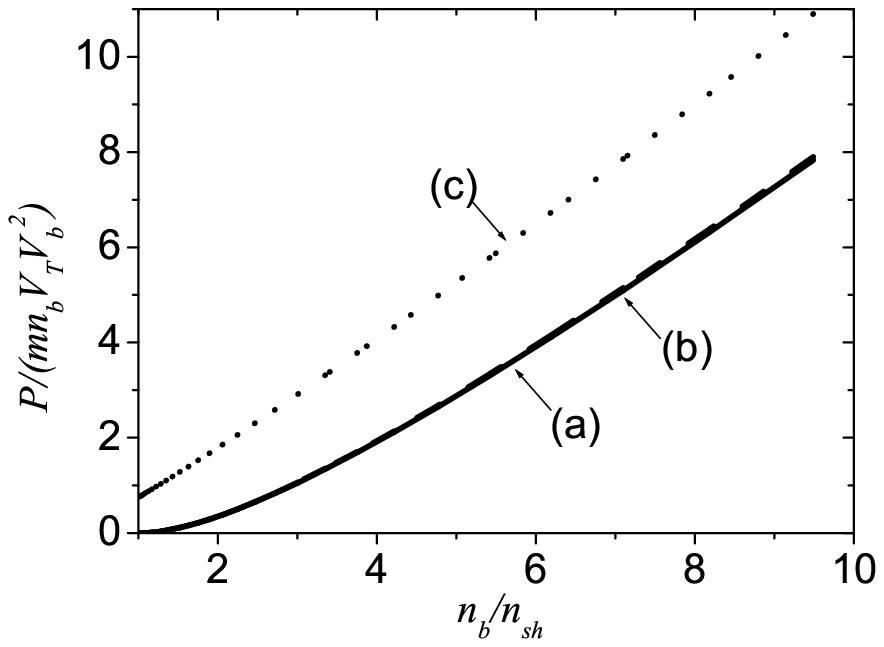} \caption{Plot of the
dimensionless power density as a function of the ratio of the bulk
plasma density to the sheath density, taking into account (a) both
$E_{1}(x)$ and $E_{b}$- solid line; (b) only $E_{b}$- dashed line;
and (c) no electric field - dotted line. }
 \label{fig4}
\end{figure}

\end{document}